\documentclass[twocolumn,english,prl,longbibliography]{revtex4-1}

\usepackage[active]{srcltx}
\usepackage{color}
\usepackage{babel}
\usepackage{amsmath}
\usepackage{amssymb}
\usepackage{graphicx}
\usepackage[unicode=true,pdfusetitle,
 bookmarks=false,
 breaklinks=true,pdfborder={0 0 0},pdfborderstyle={},backref=false,colorlinks=true]
 {hyperref}
\hypersetup{
 colorlinks,linkcolor=red,citecolor=blue}
\usepackage{breakurl}
\usepackage{lipsum}
\usepackage{dcolumn}% Align table columns on decimal point
\usepackage[T1]{fontenc}

\makeatletter

\usepackage{amsfonts}
\makeatother

\begin{document}

\preprint{APS/123-QED}

\title{Inverse Faraday Effect in an Optomagnonic Waveguide}% Force line breaks with \\

\author{Na Zhu, Xufeng Zhang, Xu Han, Chang-Ling Zou, and Hong X. Tang}
\email{hong.tang@yale.edu}

\address{Department of Electrical Engineering, Yale University, New Haven, Connecticut 06520, USA}

\date{\today}

\begin{abstract}
Single-mode high-index-contrast waveguides have been ubiquitously exploited in optical, microwave, and phononic structures for achieving enhanced wave-matter interactions. Although  micro-scale optomechanical and electro-optical devices have been widely studied, optomagnonic devices remain a grand challenge at the microscale. Here, we introduce a planar optomagnonic waveguide platform based on a ferrimagnetic insulator that simultaneously supports single transverse mode of spin waves (magnons) and highly confined optical modes. The co-localization of spin and light waves gives rise to enhanced inverse Faraday effect, and as a result, magnons are excited by an effective magnetic field generated by interacting optical photons. Moreover, the strongly enhanced optomagnonic interaction allows us to observe such effect using low-power (milliwatt level) light signals in the continuous-wave form, as opposed to high-intensity (megawatt peak power) light pulses that are typically required in magnetic bulk materials or thin films. The optically-driven magnons are detected electrically with preserved phase coherence, showing the feasibility for launching spin waves with low-power continuous optical fields. 
\end{abstract}

\maketitle
\section{Introduction}
Magnonics, the study of using spin waves as the information carrier \cite{lachance2019hybrid,kimel2019writing}, has become an emerging research field owing to recent progresses on advancing their coupling to other excitations including microwave photons \cite{zhang2014strongly,zhang2015magnon,li2019strong,harder2018level}, phonons \cite{zhang2016cavity,matthews1962acoustic,woods2001magnon,holanda2018detecting}, and optical photons \cite{hisatomi2016bidirectional,osada2018brillouin,haigh2016triple,haigh2018selection,kusminskiy2016coupled,zhang2016optomagnonic,osada2016cavity,kimel2019writing}. Compared with its phononic and electronic counterparts which typically operate at the fixed frequencies, magnonic devices possess the advantage of large frequency tunability from megahertz to hundreds of gigahertz \cite{gundougan2015solid,jobez2014cavity}. Such devices have been demonstrated in a broad range of applications, including microwave memory \cite{zhang2015magnon}, non-reciprocal devices \cite{wang2019nonreciprocity,zhang2020broadband,zhu2020magnon}, controllable logic gates \cite{rao2019analogue,chumak2015magnon,chumak2014magnon}, and radio-frequency-to-optical transducers \cite{hisatomi2016bidirectional,zhu2020waveguide}.

 In previous demonstrations, microwave signals are typically used for exciting magnons. However, optical manipulation of spins has recently emerged as a key aspect for the further development of many magnonic applications such as high density magnetic storage technology, ultrafast magnonic devices, and optically-interconnected information processing \cite{garello2014ultrafast,li2013femtosecond,stanciu2007all,braggio2017optical}. Non-thermal excitation of spin populations has been demonstrated on various materials in the study of optical helicity-driven magnetization dynamics via the inverse Faraday effect (IFE) \cite{kimel2005ultrafast,choi2017optical,hansteen2005femtosecond,garello2014ultrafast,stanciu2007all}. By utilizing the effective magnetic field created by intense circularly or linearly polarized laser pulses, the precession of magnetizations can be controlled into new equilibrium states, thus allowing coherent control of magnons by lights \cite{satoh2012directional,kanda2011vectorial}. So far, such experiments have mainly been performed on bulk materials and un-patterned thin films with a magneto-optic pump-probe apparatus \cite{kimel2005ultrafast,li2013femtosecond,stanciu2007all}. The lack of both optical and magnonic mode confinement results in weak magneto-optical interactions, thus, requiring high-intensity ultrafast laser pulses on the timescale of the sub-picosecond to achieve optical manipulations \cite{sheng1996inverse,raja1995room,horovitz1998inverse}. The realization of such ultrafast lasers are typically bulky and expensive. At the same time, the experimental requirements for the laser peak power are typically higher than megawatt level \cite{kimel2005ultrafast,li2013femtosecond,stanciu2007all}. To lower such barriers,  optomagnonic cavities with strong confinement are key to realize low power optical access for the magnon manipulation \cite{kusminskiy2019cavity}.

Considerable progresses have been made in recent years to develop micro/nano-patterned magnonic structures for efficient manipulation of magnons in the microwave domain \cite{chumak2014magnon,zhu2017patterned,krysztofik2017ultra,heyroth2019monocrystalline}, however, their applications in the optical domain are lacking. On the other hand, optical manipulation and detection of magnons has been limited in bulky structures such as highly-polished bulk spheres of yttrium iron garnet (YIG)  \cite{hisatomi2016bidirectional,osada2018brillouin,haigh2016triple,zhang2016optomagnonic,osada2016cavity} and un-patterned film surfaces \cite{demokritov2001brillouin,sebastian2015micro,demokritov2007micro,baba2019optical}, while the explorations for the micro-patterned optomagnonic devices are still missing. So far, the achieved magneto-optical interactions in existing studies remain weak because it is challenging to simultaneously fulfill both of the following two factors: the geometry engineering to localize both optical and magnonic energies in a small interaction volume with large field overlap, and the resonance enhancement to boost up the effective light-magnon interaction length.

In this work, we present integrated optomagnonic waveguides with enhanced coupling between optical lights and magnons and demonstrate IFE under continuous-wave (cw) operation. The optomagnonic waveguides are nanofabricated in a single crystalline ferrimagnetic yttrium iron garnet on gadolinium gallium garnet (YIG-on-GGG) thin film, which support single transverse mode of magnons. At the same time, the waveguides enable strong confinement of both optical and magnonic energies with orders of magnitude reduction of mode volume compared to previously demonstrated YIG devices. Moreover, the waveguides support low-loss Fabry-P\'erot cavity modes of both optical and magnonic resonances along the longitudinal direction, boosting up the effective interaction length between light and magnons. Resulting from the enhanced magneto-optical interaction, generation of gigahertz-frequency coherent magnons are realized by low-power ($\sim$1 mW) linearly-polarized cw lights via IFE. The optically generated magnons can be directly picked up with an \emph{rf} antenna while preserving the phase coherence. Our device offers the advantage of reduced power requirement for optical control of magnonic devices, and provides a chip-scale, low-cost platform for magnon-based signal processing \cite{lachance2019hybrid}.

%%%%%%%%% figure 1 %%%%%%%%%%%%%%
\begin{figure*}[t]
\centering
\includegraphics[width= 142mm]{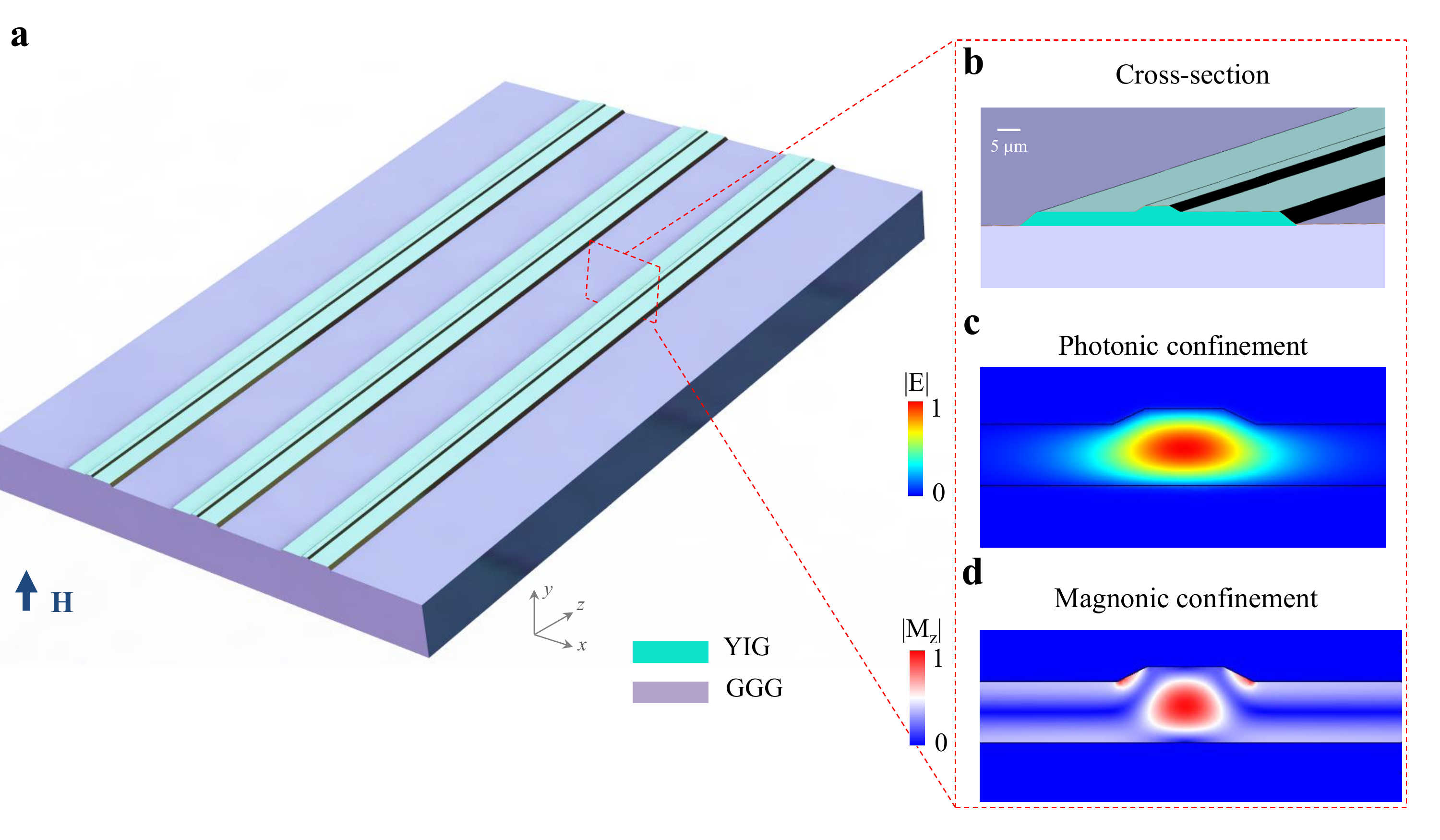}
\caption{Integrated optomagnonic waveguide. \textbf{a}. The schematic of the integrated optomagnonic waveguide. The single crystalline YIG ridge waveguides are patterned on the GGG substrate, which simultaneously confine optical photons and magnons. \textbf{b}. A schematic cross-section of YIG-on-GGG ridge waveguide. \textbf{c}. The simulated normalized electric field distribution of the optical mode. \textbf{d}. The simulated normalized magnetic field distribution of the magnon mode at the cross-section view, indicating the confinement of magnons in the YIG-on-GGG ridge waveguide.
}
\label{figure1}
\end{figure*}

%%%%%%%%%%%%%%%%%%%%%%%%%%%%%%%%%%%%%%%%%%
\section{Optomagnonic Waveguides}
%%%%%%%%%%%%%%%%%%%%%%%%%%%%%%%%%%%%%%%%%%
The single-transverse-mode magnonic waveguide is at the core of an optomagnonic circuit to confine and route magnons. To induce strong light-magnon coupling, such waveguide structures should also have low optical losses and high optical index contrast to localize the optical photons in the same geometry, forming the photonic waveguide. Additionally, to achieve direct microwave readout of the magnon signals in the waveguide, the optomagnonic circuit is required to have low microwave absorption as well.

%%%%%%%%% figure 2 %%%%%%%%%%%%%%
\begin{figure*}[htbp]
\centering
\includegraphics[width= 96mm]{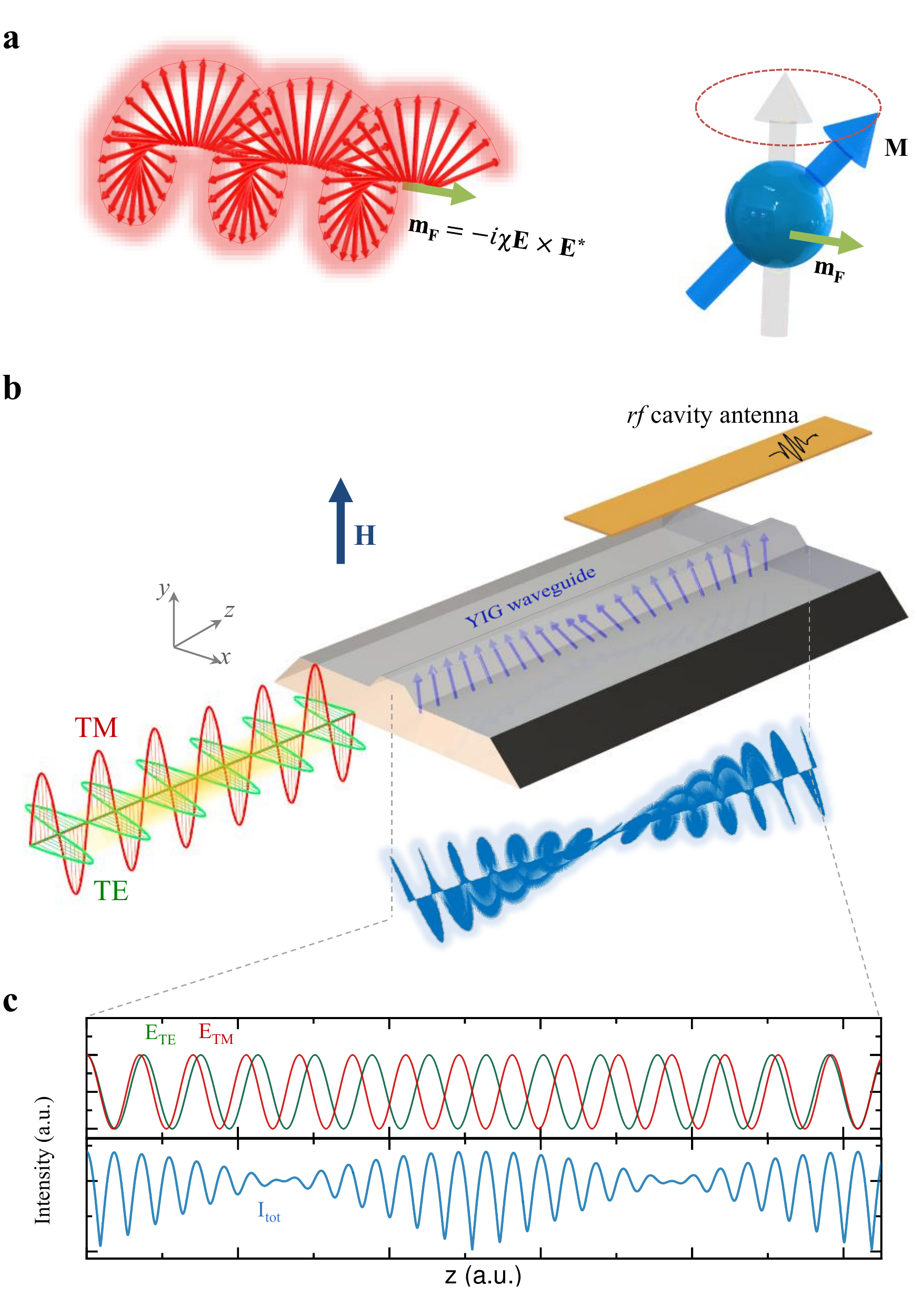}
\caption{Light-magnon interaction in a planar waveguide. \textbf{a}. {Inverse Faraday effect} in a bulk magnetic material: a circularly polarized light pulse causes the magnetization to process at a tilt angle. \textbf{b}. Interacting light and spin waves in an optomagnonic ridge waveguide. Two coherent linearly-polarized cw lights (TE \& TM) are sent to the device via the cleaved optical fibers, forming the effective circularly-polarized light when propagating along the waveguide as a result of the polarization beating, which creates an oscillating fictitious magnetic field along the $z$ axis and excites magnons. The magnon signals are read out electrically by coupling to a \textit{rf} cavity antenna. \textbf{c}. Illustration of the spatial distribution of the beat note between two coherent and orthogonally polarized cw lights. The top panel represents the electric field distribution along with $z$ axis when TE and TM standing wave lights have same intensity but slightly different effective refractive indices, and the bottom panel shows the total intensity change as a relation of the propagating length.
}
\label{figure2}
\end{figure*}
%%%%%%%%%%%%%%%%%%%%%%%%%%%

%
Based on these criteria, the single crystalline YIG-on-GGG thin film stands out as a promising candidate for building such optomagnonic waveguides. First, YIG has low optical loss (0.13 dB/cm) \cite{wood1967effect} and high optical refractive index ($\mathrm{n_{YIG}}$ $=$ 2.19) at telecom c-band. Therefore, it has a high refractive index contrast with respect to the GGG substrate ($\mathrm{n_{GGG}}$ $=$ 1.94) and air ($\mathrm{n_{air}}$ $=$ 1), allowing strong confinement for the optical field. Second, single crystalline YIG has low Gilbert damping factor ($\alpha_\textrm{YIG} \sim$ $3\times10^{-5}$) \cite{sun2012growth, sparks1964ferromagnetic} with the ferromagnetic resonance linewidth on the level of several megahertz \cite{zhang2014strongly,zhang2015magnon}. Additionally, YIG is an insulator with minimal absorption for microwave signals, which drastically eases the integration with microwave readout circuits.

Figures \ref{figure1}(a) \& (b) illustrate the optomagnonic waveguide design. An array of YIG ridge waveguides is patterned on the GGG substrate. The cross-section view shows that the YIG ridge waveguide consists of two layers: the narrow strip layer on the top and the wide slab layer at the bottom. Such a structure is designed for the following reasons. First, both transverse optical and magnonic modes are supported in such a dual layer ridge waveguide. The transverse magnon mode has the maximal field intensity at the center, dramatically boosting up the magneto-optical coupling strength by maximizing the field overlap among magnonic and optical modes. Second, the effective refractive indices of the two fundamental optical modes within such a structure are very similar to each other, resulting into cavity-enhanced IFE when both optical input lights are on-resonance. Additionally, the dual layer waveguide confines the optical field of the transverse modes mainly at the center of the bottom slab layer, reducing the scattering loss introduced by sidewall etching. 

Both optical and magnonic mode profiles are simulated using finite element method via COMSOL Multiphysics \cite{rychly2015magnonic,mruczkiewicz2014observation}, as shown in the Fig. \ref{figure1}(c) \& (d). For the magnon mode, the lateral confinement is achieved by harnessing the magnonic-index-contrast in the ridge waveguide. The fundamental transverse magnon mode is effectively a half-$\lambda$ resonance with the maximal intensity at the center of the cross section. The center YIG strip at the top layer induces additional magnonic-index contrast, leading to further localizations for the magnon mode. In particular, only one transverse magnon mode can be excited and read out electrically in our experimental configuration, because higher order transverse modes with even larger wave vector $k$ is out of the excitation and readout bandwidth ($\sim2\pi$ $\times$ 250 MHz) of the resonant antenna.

For the optical mode, only the fundamental transverse-electric (TE) and transverse-magnetic (TM) modes are supported and well-localized at the center of the etched ridge due to optical-index-contrast. In a Faraday active material, the magneto-optical interaction strength is written as \cite{kusminskiy2019cavity}
\begin{equation}
   G_{\mathrm{TE,TM},\gamma} = -\frac{i\theta_{\mathrm{F}}\lambda_{\mathrm{n}}}{4\pi}\frac{\varepsilon_{\mathrm{o}}\varepsilon_{\mathrm{r}}}{2\hbar}\int{d\mathrm{\mathbf{r}}} \cdot \mathrm{\delta\mathbf{m_{\gamma}(r)}} \cdot [\mathrm{\mathbf{E}^*_{\mathrm{TE}}(\mathbf{r})}\times{\mathbf{E}_{\mathrm{TM}}(\mathbf{r})}],
\label{equationm1}
\end{equation}
where $\delta\mathrm{\mathbf{m_{\gamma}(r)}}$ and $\mathrm{\mathbf{E}_{\mathrm{TE(TM)}}}$ are the quantized magnon magnetization and optical electric fields, respectively. $\theta_{\textrm{F}}$ is the Faraday rotation coefficient, $\lambda_{\mathrm{n}}$ $=$ $\lambda_{\mathrm{o}}/n$ with $n$ being the optical refractive index and $\lambda_{\mathrm{o}}$ being the vacuum wavelength, and  $\epsilon_{\mathrm{r}}$ is the material relative permittivity (Detailed derivation can be found in Appendix A).

The magneto-optical coupling strength is limited by a suppression factor $\eta$ among the magnon, optical TE, and optical TM modes, proportional to the integral in the Eq.\,(\ref{equationm1}). To quantify this effect, we defined the $\mathrm{V_M}$, $\mathrm{V_{TE}}$, and $\mathrm{V_{TM}}$ as the mode volume of the magnon, optical TE, optical TM modes, respectively. (Detailed derivations can be found in Appendix A.) At the same time, the overlap volume between magnonic and optical modes are defined as $V_{\mathrm{MO}}$ ($V_{\mathrm{MO}} < \mathrm{min\{\mathrm{V_{M}} ,\mathrm{V_{TE}}, \mathrm{V_{TM}\}}}$, Appendix A). 

\begin{equation}
    \eta \approx \frac{V_{\mathrm{MO}}}{\sqrt{{{V_\mathrm{M}}{{V_\mathrm{TE}}}}{{V_\mathrm{TM}}}}}.
\label{eq_fo}    
\end{equation}
From Eq. (\ref{eq_fo}), we see that the MO coupling is limited by the smallest achievable mode volume for all three participating modes. In this device configuration, we have $V_{\mathrm{TE}} \approx V_{\mathrm{TM}} \geq V_{\mathrm{MO}}$, the coupling strength is approximately proportional to $1/\sqrt{V_{\mathrm{M}}}$, favoring the smaller magnonic mode volume. Thus, the highly confined magnonic transverse mode greatly reduces the $\mathrm{V_M}$ by around three orders of magnitude compared with the YIG sphere experiments utilizing  uniform Kittel modes \cite{zhang2016optomagnonic, osada2016cavity}. Both optical and magnonic fields are strongly localized in a small volume with maximal overlap, desired for the strong magneto-optical interactions and optical generation of gigahertz-frequency magnons via {IFE} \cite{kusminskiy2019cavity}.

%%%%%%%%% figure 3 %%%%%%%%%%%%%%
\begin{figure*}[htbp]
\centering
\includegraphics[width= 145mm]{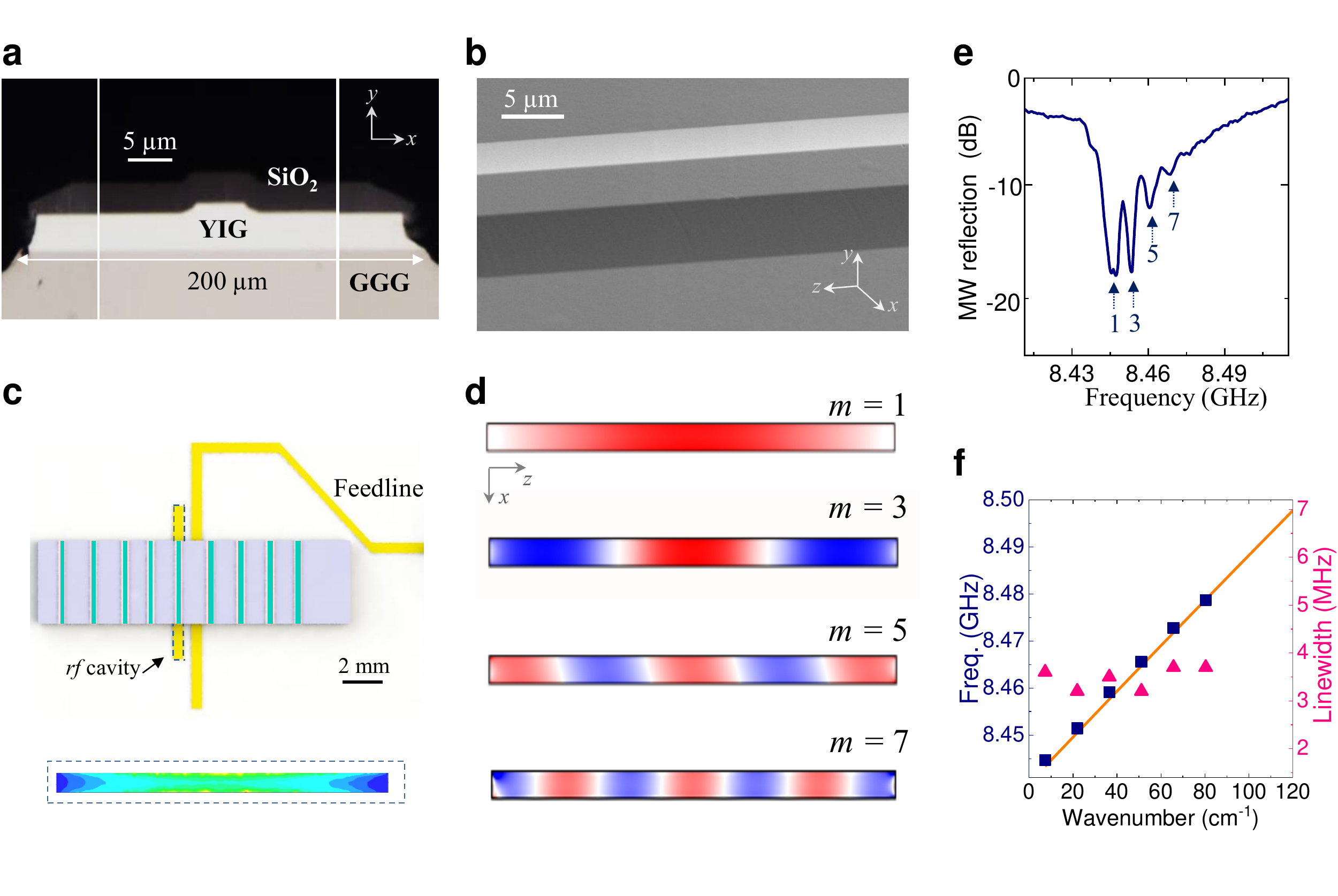}
\caption{Characterization of the magnonic waveguide. \textbf{a}. The cross-sectional view of the ridge waveguide. \textbf{b}. The top-view scanning electron microscope (SEM) image of the magnonic waveguide (zoomed-in at the center of the ridge waveguide). \textbf{c}. Schematic of the coplanar \emph{rf} circuit for electrical excitation and readout of magnons. The yellow area is the copper layer, which also covers the backside of the circuit, with the rest white area being the high-dielectric constant substrate. The microstrip cavity (dashed box) supports a microwave half-$\lambda$ resonant mode, which can effectively couple with magnon modes. The bottom plot in the dashed box shows the magnetic field distribution of the \emph{rf} cavity half-$\lambda$ microwave mode resonates at $\sim$2$\pi \times$8.45 GHz, showing the maximized \emph{rf} magnetic field intensity at the center of the \textit{rf} cavity, overlapping with the magnonic waveguide. \textbf{d.} The simulated magnon modes spatial magnetization distribution, indicating the spin wave modes formed along the waveguide length direction. \textbf{e.} The microwave reflection spectrum when the waveguide is magnetized perpendicularly with $\sim$ 4700 Oe static magnetic field. Only the modes with odd mode number are measured because of the effective non-zero coupling with the microwave $rf$ cavity. \textbf{f.} Dispersion relation of the spin wave in the YIG magnonic waveguide. Blue squares and purple triangles are the experimentally extracted dispersion and magnon linewidth. Solid line is calculated dispersion. 
}
\label{figure3}
\end{figure*}
%%%%%%%%%%%%%%%%%%%%%%%%%%%

\section{Inverse Faraday Effect}

Previous studies of IFE involve the generation of an effective quasi-static axial magnetic field by circularly polarized light radiation propagating through a bulk Faraday-active material \cite{sheng1996inverse,raja1995room,horovitz1998inverse}. As illustrated in the Fig. \ref{figure2}(a), for Faraday-active materials with isotropic magneto-optical properties, optically induced magnetization is written as $\mathbf{m_{\mathrm{F}}}$ $=$ $-i{\chi}\mathbf{E} \times \mathbf{E^{*}}$ \cite{van1965optically,mikhaylovskiy2012ultrafast}, where $\mathbf{E}$ is the complex electric field of the incident light, and $\chi$ is the magneto-optical susceptibility proportional to the material Faraday rotation coefficient \cite{pershan1966theoretical}. Such optically induced axial magnetic field will cause the perturbation to magnetizations within the sample, resulting into the procession of the magnetic moments and effective excitation of spin waves. Equation \ref{equationm1} also implies the need of a non-trivial polarization of the optical field to obtain a finite magneto-optical coupling term.

As opposed to commonly used impulsive methods, our demonstration employs two phase-coherent cw lights, as illustrated in Fig. \ref{figure2}(b). The two lights with orthogonal polarizations (TE \& TM) have slightly different frequencies ($\omega_{\mathrm{TM}}$ $=$ $\omega_{\mathrm{TE}}$ $+$ $\delta{\omega}$). The device is magnetized perpendicularly via a static magnetic field $\mathrm{\mathbf{H}}$. When propagating in the waveguide, the TE and TM-polarized lights have different effective refractive indices ($\mathrm{n_{TE}}$ $\neq$ $\mathrm{n_{TM}}$), leading to the temporal beating which in turn induces axial magnetization oscillation at \textit{rf} frequencies. In the Fig. \ref{figure2}(b), the blue curve shows the trace of the total electric vector when the waveguide length is set to be one polarization beat length $l_{\mathrm{{beat}}}$ $=$ $\frac{\lambda}{|\mathrm{n_{TE}-n_{TM}}|}$, demonstrating the effective elliptically-polarized light formed in the waveguide due to the polarization spatial beating \cite{huan2000realization,shevchenko2019interference}. Correspondingly, in the Fig. \ref{figure2}(c), the upper panel plots the normalized individual electric field distributions of TE and TM standing waves with different effective refractive indices as a relation of the propagation length within the waveguide. The bottom panel shows the total optical intensity change as a relation of the propagation length.  Note that the generation of the optically induced \textit{rf} magnetization does not require the length of the waveguide to be longer than the $l_{\mathrm{{beat}}}$. The $\mathrm{\mathbf{m_{F}}}$ is along the $z$ axis with $\mathrm{\mathbf{m_{F}}}(x,y,z)$ $=$ $\frac{V}{2{\pi}c}\frac{\lambda}{n}\sqrt{I_\mathrm{{TE}}(x,y)I_\mathrm{{TM}}(x,y)}\mathrm{cos}[(\omega_{\mathrm{TE}}-\omega_{\mathrm{TM}})t - \frac{2\pi}{\lambda}(n_\mathrm{{TE}}-n_\mathrm{{TM}})z] {\widehat{\mathbf{z}}}$, where $V$, n, $\lambda$ are the Verdet constant, material refractive index, operating wavelength, respectively \cite{raja1995room,van1965optically}. $I_\mathrm{{TE(TM)}}(x,y)$ is the optical intensity which has the spatial dependence, and $n_\mathrm{{TE(TM)}}$ is the effective refractive index of the TE(TM) light in the waveguide. It can be seen that $\mathbf{m}_\mathbf{{F}}$ oscillates at the frequency difference between TE and TM lights (${\omega_{\mathrm{beat}}} = \omega_{\mathrm{TE}} - \omega_{\mathrm{TM}}$), which perturbs the magnetic moments aligned along the $y$ axis.

When the frequency of the optically induced oscillating magnetization matches the magnon mode frequencies ($\omega_{\mathrm{beat}}$ $\approx$ $\omega_{\gamma}$), spin waves are excited via IFE. Instead of using an optical probe as in typical impulsive implementations \cite{kimel2005ultrafast,PhysRevB.78.104301}, here we design a \textit{rf} cavity antenna which can couple with the selected magnon modes in the YIG waveguide and read them out in the microwave domain with high extraction efficiency. 

Figures \ref{figure3}(a) \& (b) show the optical micrograph and scanning electron micrograph (SEM) images of the fabricated YIG ridge waveguide at the cross-section view and top view, respectively. These waveguide structures are patterned based on our fabrication techniques that not only are compatible with the standard semiconductor fabrication processes but also guarantee excellent surface finish and controllable optical dispersion (see Appendix B). The SEM image is taken when zoomed-in at the center 1.2-$\mu$m-tall confinement step, showing a very smooth surface finish which minimizes the optical loss. The device is diced with a length of 4.2 mm, and careful surface polishing is applied for both facets to minimize both optical and magnonic scattering losses. Before studying the magneto-optical interaction, we first characterize the waveguide magnon resonances via the microwave reflection measurement. As shown in  Fig. \ref{figure3}(c), a \textit{rf} coplanar microstrip cavity is used to couple with the waveguide magnonic modes. The \textit{rf} microstrip antenna is patterned on the PCB circuit and supports a half-$\lambda$ microwave mode resonates at $\sim$ $2\pi$ $\times$ 8.444 GHz with the external coupling rate and intrinsic decay rate being $\kappa_{\mathrm{e,e}}/2\pi$ $=$ 165 MHz and $\kappa_{\mathrm{e,i}}/2\pi$ $=$ 85 MHz, respectively. Such microwave mode has maximized \textit{rf} magnetic field at the center of the microstrip cavity, which facilitates efficient excitation of the magnon modes when the waveguide is aligned and positioned at the center of the \textit{rf} microstrip. 

\section{Magnonic Modes Characterization}

While the magnonic waveguide is designed to support a single mode in the transverse direction, magnons are allowed to propagate along the waveguide longitudinal direction to form standing waves bounded by the waveguide facets. To further understand the longitudinal magnon modes within the magnonic waveguide, we perform finite-element simulations to derive the spatial dependence of the mode field distributions using COMSOL Multiphysics \cite{mruczkiewicz2013standing,mruczkiewicz2014observation,rychly2015magnonic}. Here, the waveguide is magnetized perpendicularly, thus, forward volume magnetostatic waves (FVMSWs) are supported in the thin film configuration \cite{stancil2009spin,chen2018spin}. Due to the geometrical confinement, standing wave modes are formed along both transverse ($x$ axis) and longitudinal ($z$ axis) directions. With the length of the waveguide being approximately 4.2 mm, for the longitudinal confinement, multiple standing wave magnon modes can be formed by FVMSWs with \emph{free spectra range} (FSR) around $2\pi$ $\times$ 7.3 MHz. Field distributions for the modes with odd mode numbers are simulated and plotted in the Fig. \ref{figure3}(d) (mode number $m$ $=$ 1, 3, 5, ...). Noticeably, only odd modes can have a finite coupling strength with the \textit{rf} antenna during the microwave reflection measurement and readout of the optically-generated magnon signals, as the even modes have cancelled coupling strengths with the antenna microwave cavity field.

%%%%%%%%% figure 4 %%%%%%%%%%%%%%
\begin{figure}[htbp]
\centering
\includegraphics[width= \linewidth]{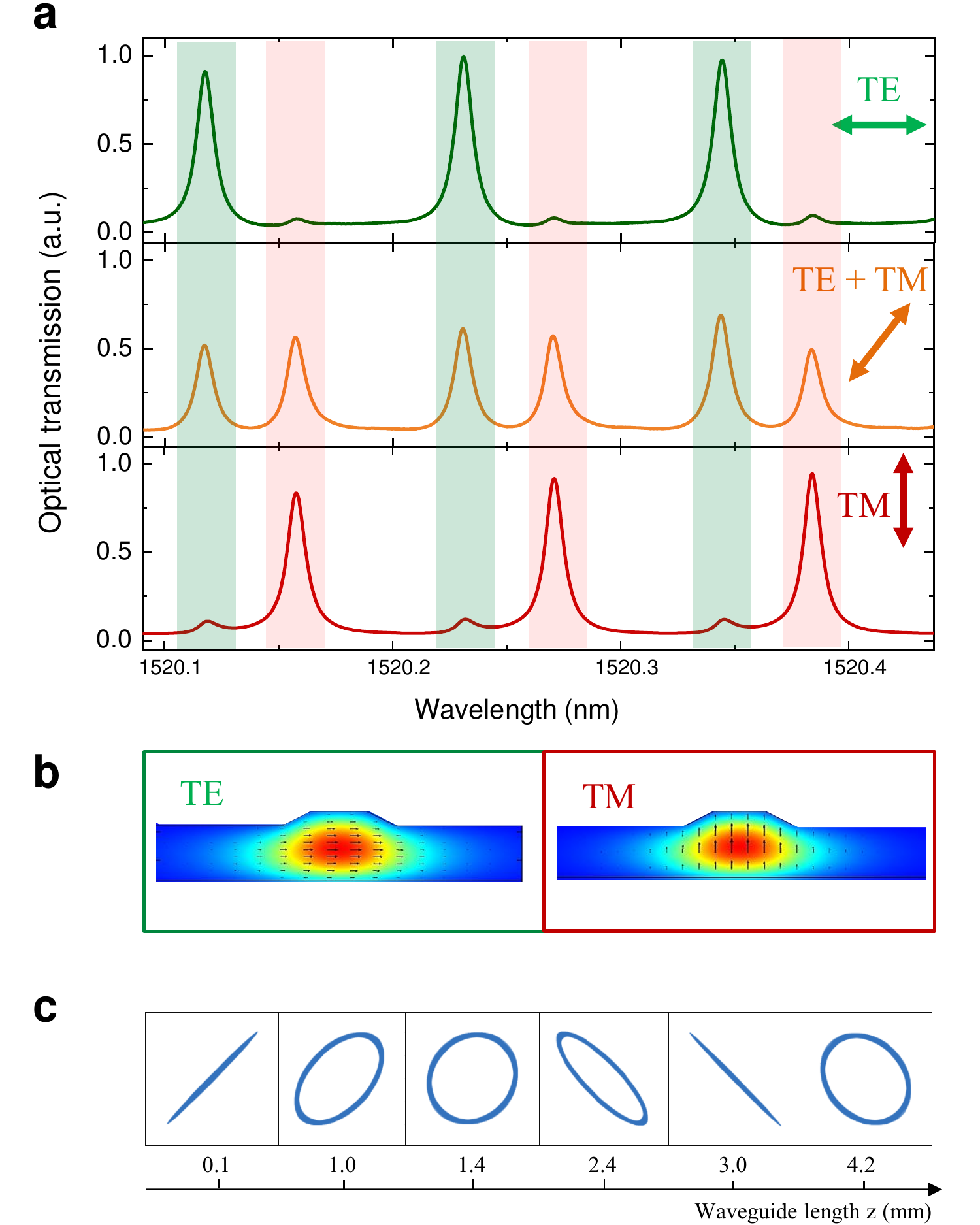}
\caption{Experimental characterization of the photonic waveguide. \textbf{a}. The optical transmission spectra of the waveguide are experimentally measured when the polarization of the input cw lights is aligned at TE, (50\% TE $+$ 50\% TM), TM, respectively. \textbf{b}. The simulated electrical field distributions of the TE and TM polarized fundamental modes, showing very similar mode profiles ($n_{\mathrm{TE}} = 2.1829$, $n_{\mathrm{TM}} = 2.1827$, $l_{\mathrm{beat}} = 7.6$ mm). \textbf{c}. The calculated electric field vector of optical mode at the $x-y$ plane when propagating along the waveguide longitudinal direction, showing the pattern of elliptically-polarized light due to the polarization spatial beating between TE and TM light. Each plot considers 200 $\mu$m propagation distance.}
\label{figure4}
\end{figure}
%%%%%%%%%%%%%%%%%%%%%%%%%%%

%%%%%%%%% figure 5 %%%%%%%%%%%%%%
\begin{figure*}[htbp]
\centering
\includegraphics[width= 151mm]{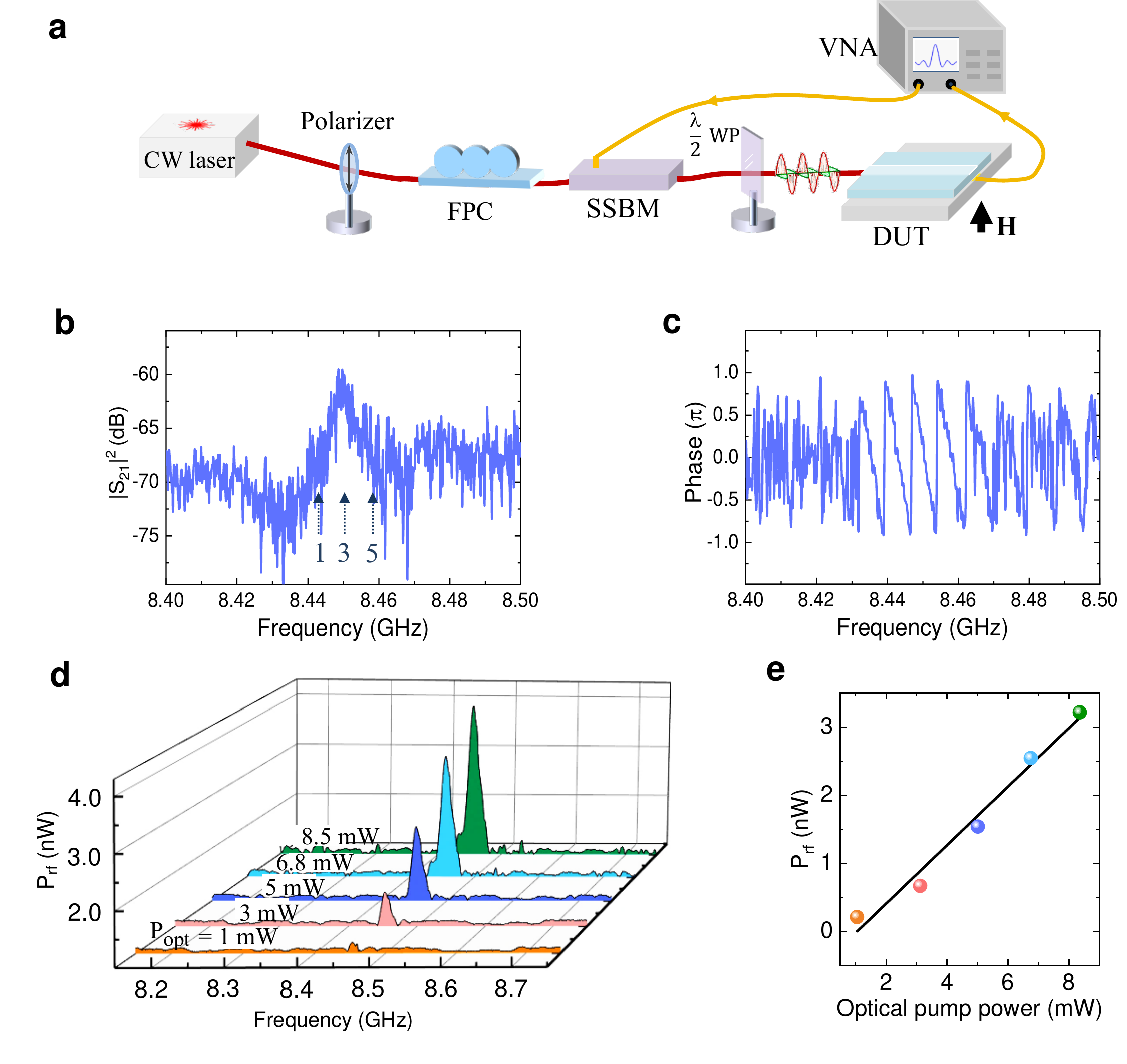}
\caption{Readout of the {inverse Faraday effect} generated magnons. \textbf{a}. The experimental setup. (FPC: fiber polarization controller; SSBM: single sideband modulator; WP: waveplate; DUT: device under test; VNA: vector network analyzer.) \textbf{b \& c}. The magnitude and phase of the measured optically generated magnon signals. (Input local oscillator optical power: 5 mW, VNA output power: 1 mW) \textbf{d}. The optical pump power dependence of the optically generated magnon measured via the $rf$ cavity antenna \textbf{e}. Linear optical power dependence,  measured (scatter) and fitted (line). }
\label{figure5}
\end{figure*}

%%%%%%%%%%%%%%%%%%%%%%%%%%% 

The microwave reflection spectrum is measured by tuning the magnon resonances within a coupling bandwidth of \textit{rf} antenna ($2\pi$ $\times$ 250 MHz). From Fig. \ref{figure3}(e), four magnon resonances are clearly observed with different $k_{{l}}$ vectors in the longitudinal direction ($k_{{l}}$ $=$ $m\pi/l$). The magnon dispersion relation is extracted and plotted in the Fig. \ref{figure3}(d). Here, the theoretical dispersion relation is calculated for the FVMSW using equation $\omega = \sqrt{{\omega_{\mathrm{o}}}\big[{\omega_{\mathrm{o}}} + {\omega_{\mathrm{m}}} (1 - \frac{1 - e^{-k_{l}d}}{k_{l}d}) \big]}$, where $\omega_{\mathrm{o}} = {\gamma}(H - H_{\mathrm{d}})$, $\omega_{\mathrm{m}} = {\gamma}4{\pi}M_{\mathrm{s}}$ \cite{stancil2009spin,zhang2016superstrong}, and $\gamma = 2\pi \times 2.8$ MHz/Oe is the gyromagnetic ratio. $H$ is the static bias magnetic field, $H_{\mathrm{d}} \approx 1750$ Oe is the demagnetizing field of YIG, and $4{\pi}M_{\mathrm{s}} = 1750$ G is the saturation magnetization of single crystalline YIG. $d$ is the waveguide thickness.

\section{Optical Modes Characterization}

Next, we characterize the photonic performance of the optomagnonic waveguide. The ridge waveguide geometry localizes the optical fields at the center of the etched strip, and supports fundamental  transverse TE \& TM modes. The waveguide facets are polished to ensure low optical loss and then deposited with high reflection metallic mirrors, forming a waveguide Fabry-P\'erot cavity that produces a series of TE \& TM resonances. These two types of modes have similar cross-sectional intensity distributions, as shown in the Fig. \ref{figure4}(b), resulting in slightly different effective refractive indices and FSRs.

In Fig. \ref{figure4}(a), the optical transmission spectra are measured by rotating the polarization of the input cw light. The results clearly show that the TE and TM modes resonant at slightly different frequencies. At around 1520.23 nm, the frequency difference between the adjacent TE and TM modes is measured to be $\sim$ $2\pi$ $\times$ 8.45 GHz, matching the magnon resonant frequencies. With the effective refractive indices difference, when coherent TE and TM lights are sent to these two adjacent modes,  polarization beating occurs while propagating in the waveguide and then generates the cavity-enhanced optically-driven \textit{rf} magnetization. In Fig. \ref{figure4}(c), based on the simulated effective refractive indices, the electric field vector is plotted in the $x-y$ plane as a relation of the propagation distance, when the coherent TE \& TM lights of equal intensity oscillating at 1520.23 nm and 1520.16 nm, respectively, are launched into the waveguide. The calculated result shows the evolution of the input light from linearly-polarized state to elliptically-polarized state due to the beating.

\section{Optical Excitation of Magnons}

Finally, we demonstrate the generation of magnons by coherent optical lights, and incorporate the efficient microwave frequency readout of the generated magnon signals. The experimental setup is shown in the Fig. \ref{figure5}(a). The waveguide device is biased via a static magnetic field in the out-of-plane configuration, with magnon modes corresponding to the measurement results shown in the Fig. \ref{figure3}(e) \& (f). A cw laser light source at 1520.23 nm (TE mode resonant wavelength) is used to generate the coherent TE and TM lights. After passing through optical filters, an optical polarizer is used to ensure the linear polarization of the input light. 

In order to use two coherent lights with tunable frequency differences to generate \textit{rf} optically-induced magnetization, a single-sideband modulator (SSBM) is used in our experiment. The local oscillator (pump light from the laser diode) and the modulated single sideband have the identical polarization after the modulator, resulting into net zero $m_{\mathrm{F}}$. Thus, we use a half-$\lambda$ waveplate to rotate the polarization axis of both local oscillator and modulated sideband to $\pi/4$ relative to both TE and TM electric field directions, leading to half-TE and half-TM linear polarization for both lights. The YIG waveguide is inherently an excellent Fabry-P\'erot filter with polarization selection capabilities. Therefore, the local oscillator is set at the frequency of waveguide TE resonant mode ($\omega_{\mathrm{TE}}$), and the SSBM is modulated at the frequency $\delta{\omega}$. Once $\delta{\omega}$ is tuned to equal the frequency difference between the TE mode and its adjacent TM mode, the modulated sideband frequency will match the waveguide TM mode resonant frequency ($\omega_{\mathrm{TM}}$ $=$ $\omega_{\mathrm{TE}}$ $+$ $\delta{\omega}$, $\delta{\omega}$ $=$ $\omega_{\mathrm{beat}}$). Due to the filtering effect of the Fabry-P\'erot waveguide modes, only the TE-component of the local oscillator and the TM-polarized sideband lights can efficiently couple into the waveguide, successfully generating the \textit{rf} magnetization $m_{\mathrm{F}}$. When the output modulation frequency $\delta{\omega}$ from the VNA matches the magnon mode frequencies, the magnons can be excited in the waveguide and couple to the \textit{rf} antenna, then fed into the VNA to obtain the transmission spectra.

Owing to the strong mode confinement and large field overlap between magnons and photons, the magneto-optical coupling strength is greatly enhanced compared to the previous studies in  bulk YIG samples \cite{hisatomi2016bidirectional,osada2016cavity,zhang2016optomagnonic}. For the lowest order magnon mode ($m=1$), the single-photon magneto-optical coupling strength $G_{\mathrm{TE,TM,1}}/2\pi$ is analytically obtained by integrating over the whole volume (Appendix A) to be 8.15 Hz. The photon-number-enhanced magneto-optical coupling strength is proportional to the square root of the intra-cavity photon number, written as $g = \sqrt{n_{\mathrm{c}}}G$. With 5 mW TE pump light, the photon-number-enhanced magneto-optical coupling strength is calculated to be $g_{\mathrm{TE,TM,1}} = 2\pi \times 10.84$ kHz. As shown in the Fig. \ref{figure5}(b), the optically-generated magnon signal is measured electrically via the VNA transmission measurement. The transmission peak corresponds to the contribution from the first three magnon modes ($m$ $=$ 1, 3, 5). For this ridge waveguide device, because the waveguide length (4.2 mm) is shorter than the polarization beat length ($\sim$ 7.6 mm), all the odd number magnon modes can be excited with non-zero coupling strength. If the waveguide length is longer than the polarization beat length, phase match conditions are required for different magnon modes, as some modes will have the cancelled coupling with the optical beat field.

Figure \ref{figure5}(d) plots the power of measured IFE-generated magnons ($\mathrm{P_{rf}}$) when optical pump power is increased. In the frequency range where magnon modes exist, $\mathrm{P_{rf}}$ increases with the optical pump power. Meanwhile, the transmission outside of the magnon resonance is at the background level. In the Fig. \ref{figure5}(e), we study the dependence of excited magnon on the input optical pump power before the optical polarizer. The linear dependence is clearly manifested and consistent with the inverse Faraday effect model.

Interestingly, the transmission spectrum does not inherit the discrete mode feature as show in the microwave reflection measurement (Fig. \ref{figure3}(c)). This can be explained by the fact that the longitudinal magnon modes are closely compacted in the frequency domain, with their FSR ($\sim2\pi$ $\times$ 7.3 MHz) similar to the resonant linewidth ($\sim2\pi$ $\times$ 3.5 MHz). At the same time, the optical mode linewidth is much wider ($\sim2\pi$ $\times$ 1 GHz). Therefore, magnons can be effectively generated in the detuned scenario $\omega_{\mathrm{beat}} \approx \omega_{\mathrm{\gamma}} \approx 2\pi \times 8.45$ GHz. Due to existence of multiple magnon modes, at certain optical beating frequency $\omega_{\mathrm{beat}}$, the generation of magnons is assisted by the adjacent modes. Such collective addition smooths the transmission spectrum, as shown in Fig. \ref{figure5}(b), without the discrete feature. This also explains why the optically excited magnon signal is peaked at the third magnon mode ($m$ = 3), instead of the fundamental mode where the largest coupling strength is observed with a \textit{rf} drive (Fig. \ref{figure3}(e)). The phase response shown in the Fig. \ref{figure5}(c) clearly maintains the coherence between the optical drive and the generated magnons.

\section{Discussion}

In conclusion, we establish a planar optomagnonic waveguide architecture based on a single crystal YIG-on-GGG material platform. Low loss single-transverse-mode magnonic and photonic waveguide has been demonstrated. The engineered YIG ridge waveguide localizes and routes both magnons and optical photons in a small interaction volume, achieving a significant improvement in magneto-optical interaction strength with respect to the state-of-the-art demonstrations \cite{osada2016cavity,hisatomi2016bidirectional,zhang2016optomagnonic,PhysRevLett.121.199901}. Such improved performance enables the coherent optical control and generation of magnons with milliwatt-level low power cw optical photons via the inverse Faraday effect. Remarkably, in this experiment, the optically excited magnons can be read out coherently through the electrical channel, linking both optical and microwave domains in a waveguide device platform. This integrated optomagnonic structure demonstrates a key building block for on-chip low-power optically controllable magnonic circuits, opening doors for many applications such as magnon-based information transducer and non-reciprocal devices.

\section*{Funding}
We acknowledge funding from National Science Foundation (EFMA-1741666). H.X.T acknowledge support from a previous DARPA MTO/MESO grant ( N66001-11-1-4114), an ARO  grant (W911NF-18-1-0020) and Packard Foundation. 

\section*{Acknowledgments}
The authors acknowledge fruitful discussions with the team members of the EFRI Newlaw program led by Ohio State University: Andrew Franson, Seth Kurfman, Denis R. Candido, Katherine E. Nygren, Yueguang Shi, Kwangyul Hu, Kristen S. Buchanan, Michael E. Flatt\'e, and Ezekiel Johnston-Halperin. The authors thank M. Power and M. Rooks for the assistance in device fabrications.

\appendix

\section*{Appendix A. Magneto-Optical Coupling Strength}

In this section, we will study the generation of magnons in the optomagnonic waveguide via the \emph{Inverse Faraday Effect}. In a Faraday active material, the system electromagnetic energy is modified by the coupling between the optical electric field and the magnetization \cite{kusminskiy2019cavity}
\begin{equation}
    U_{\mathrm{MO}} = -\frac{i}{4}f\varepsilon_{\mathrm{o}}\varepsilon_{\mathrm{r}}\int\mathrm{ d\mathbf{r}d\mathbf{M(r)} \cdot [\mathbf{E^{*}(r)}\times\mathbf{E(r)}]}.
\label{eq1}
\end{equation}
Here, the relative permittivity of YIG $\varepsilon_{\textrm{r}}$ is 4.84, $\mathbf{E(r)}$ is the electric field, and $f$ is the material property constant related to the Faraday rotation coefficient by the expression $\theta_{\textrm{F}}$ $=$ $\frac{\omega fM_{\textrm{S}}}{2c\sqrt{\varepsilon_{\textrm{r}}}}$, with $M_{\textrm{S}}$ $=$ 139,260 A/m being the saturation magnetization of YIG thin films. We can see that the cross product term is proportional to the \emph{optical spin density} $\mathrm{\mathbf{S(r)}}$ $=$ $\frac{\varepsilon}{2i\omega}[\mathrm{\mathbf{E^*(r)}}\times\mathrm{\mathbf{E(r)}}]$, thus, the non-trivial polarizations are required to achieve magneto-optical coupling.

The magnetization $\mathbf{M}$ $=$ $\mathbf{M_\mathrm{o}}$ $+$ {$\delta$}$\mathrm{\mathbf{m}}$, where $\mathbf{M_\mathrm{o}}$ is the ground state magnetization and $\delta\mathrm{\mathbf{m}}$ is the weak excitation. For small deviations $|\delta\mathrm{\mathbf{m}}|$ $\ll$ 1, we can quantize the spin wave in terms of harmonic oscillators (magnon modes), meaning that 
\begin{equation}
    \delta\mathrm{\mathbf{m}(\mathbf{r},t)} \rightarrow \frac{1}{2}\sum_{\gamma}[\delta\mathrm{\mathbf{m}_\gamma}(\mathrm{\mathbf{r}})\hat{m}_\gamma + \delta\mathrm{\mathbf{m}_\gamma^*}(\mathrm{\mathbf{r}}){\hat{m}_\gamma}^\dagger].
\label{eq2}
\end{equation}

Meanwhile, the optical electric field can be readily quantized as
\begin{equation}
    \mathrm{\mathbf{E}(\mathbf{r},t)} \rightarrow \frac{1}{2}\sum_{\beta}[\mathrm{\mathbf{E}_\beta}(\mathrm{\mathbf{r}})\hat{a}_\beta + \mathrm{\mathbf{E}_\beta^*}(\mathrm{\mathbf{r}}){\hat{a}_\beta}^\dagger].
\label{eq3}
\end{equation}

From Eq. (\ref{eq1}), the magneto-optical coupling Hamiltonian is written as 
\begin{equation}
    \hat{\mathrm{H}}_{\mathrm{MO}}/\hbar = \sum_{\alpha\beta\gamma}G_{\alpha\beta\gamma}\hat{a}_{\alpha}^\dagger\hat{a}_{\beta}\hat{m}_{\gamma} + \sum_{\alpha\beta\gamma}G^*_{\alpha\beta\gamma}\hat{a}_{\beta}^\dagger\hat{a}_{\alpha}{\hat{m}_{\gamma}}^\dagger,
\label{eq4}
\end{equation}
where ${a}_{\alpha(\beta)}$ is the annihilation operators for different optical mode, ${m}_{\gamma}$ represents different magnon modes. The magnon-optical coupling strength
\begin{equation}
    G_{\alpha\beta\gamma} = -\frac{i\theta_{\mathrm{F}}\lambda_{\mathrm{n}}}{4\pi}\frac{\varepsilon_{\mathrm{o}}\varepsilon_{\mathrm{r}}}{2\hbar}\int{d\mathrm{\mathbf{r}}}\delta\mathrm{\mathbf{m(r)}}.[\mathrm{\mathbf{E}^*_\alpha(\mathbf{r})}\times{\mathbf{E}_\beta(\mathbf{r})}],
\label{eq5}
\end{equation}
where $\lambda_{\mathrm{n}}$ $=$ $\lambda_{\mathrm{o}}/n$, $n$ $\approx$ 2.19 for YIG. 

In our experiment, we engineer the optical effective refractive indices to utilize waveguide TE and TM modes for magnon excitations. Here, we denote $\hat{a}_{\alpha}$ $\rightarrow$ $\hat{a}_{\mathrm{TE}}$ and $\hat{a}_{\beta}$ $\rightarrow$ $\hat{a}_{\mathrm{TM}}$. The effective mode volume for the optical modes is commonly written as
\begin{equation}
    \mathrm{V^{TE(TM)}_E} = \frac{\int{d\mathrm{\mathbf{r}^3}}{|\mathbf{E}_{\mathrm{TE(TM)}}(\mathrm{\mathbf{r}})|^2}}{\mathrm{max}{\mathrm\{|\mathbf{E}_{\mathrm{TE(TM)}}}(\mathrm{\mathbf{r}})|^2\}}.
\label{eq6}
\end{equation}
Similarly, the magnon effective mode volume is defined as 
\begin{equation}
    \mathrm{V^{\gamma}_M} = \frac{\int{d\mathrm{\mathbf{r}^3}}{|\mathbf{\delta\mathrm{\mathbf{m}_\gamma}}(\mathrm{\mathbf{r}})|^2}}{\mathrm{max}{\mathrm\{{|\mathbf{\delta\mathrm{\mathbf{m}_\gamma}}(\mathrm{\mathbf{r}})|^2}}\}}.
\label{eq7}
\end{equation}

After fields normalization, the single photon magneto-optical coupling strength is reduced to
\begin{equation}
\begin{split}
    G_{\mathrm{TE,TM},\gamma} = -\frac{i\theta_{\mathrm{F}}\lambda_{\mathrm{n}}}{4\pi}\frac{1}{2}\sqrt{\frac{g\mu_{\mathrm{B}}}{{M_S}}}\frac{\sqrt{\omega_{\mathrm{TE}}\omega_{\mathrm{TM}}}}{\sqrt{{{V_\mathrm{M}^{\gamma}}}{{V_\mathrm{E}^{\mathrm{TE}}}}{{V_\mathrm{E}^{\mathrm{TM}}}}}} \\
    \times \int\frac{{d\mathrm{\mathbf{r}}}\delta\mathrm{\mathbf{m(r)}}.[\mathrm{\mathbf{E}^*_{\mathrm{TE}}(\mathbf{r})}\times{\mathbf{E}_{\mathrm{TM}}(\mathbf{r})}]}{{\mathrm{max}{\mathrm\{{|\mathbf{\delta\mathrm{\mathbf{m}_\gamma}}(\mathrm{\mathbf{r}})}}\}}.{{\mathrm{max}\mathrm\{|\mathbf{E}_{\mathrm{TE}}}(\mathrm{\mathbf{r}})\}}.{{\mathrm{max}\mathrm\{|\mathbf{E}_{\mathrm{TM}}}(\mathrm{\mathbf{r}})|\}}},
\label{eq8}
\end{split}
\end{equation}
where $g$ is the Land\'e factor and $\mu_{\mathrm{B}}$ is Bohr magneton. The integration represents the mode overlap among optical and magnon modes and is defined as $\mathrm{V_{om}}$ in Eq. (\ref{eq_fo}).

%%%%%%%%% figure S1 %%%%%%%%%%%%%%
\begin{figure}[htbp]
\centering
\includegraphics[width= \linewidth]{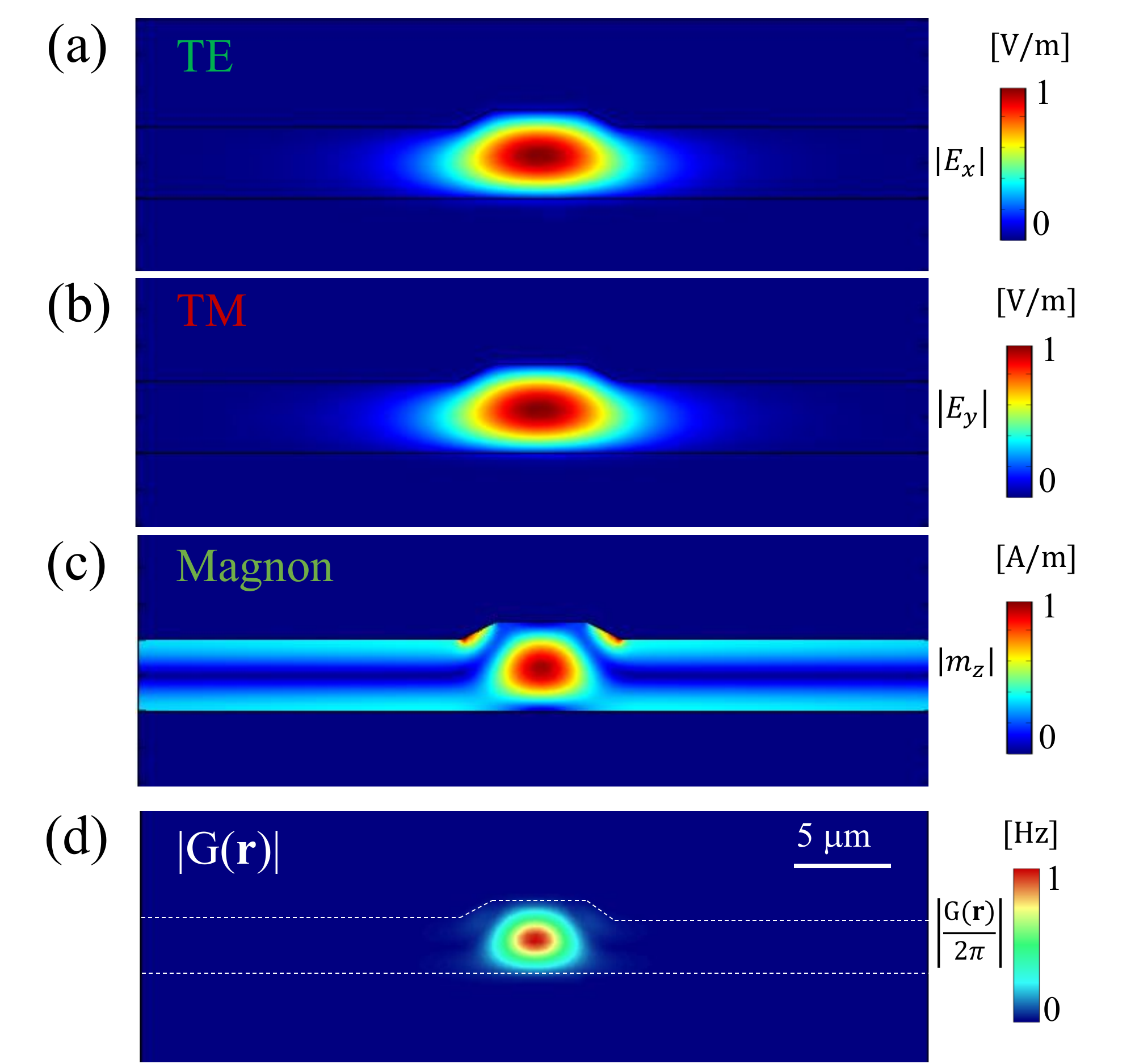}
\caption{(a) \& (b) plots the field distributions for the optical TE and TM modes. (c) is the mode profile of the magnon modes at the same cross section view. (d) shows the spatial distribution of the magneto-optical coupling strength. $|G(\mathrm{\mathbf{r}})|$ is calculated from Eq. (\ref{eq8}) then normalized.}
\label{figures1}
\end{figure}
%%%%%%%%%%%%%%%%%%%%%%%%%%%

In our experimental configuration, the magnon excitation process involves both TE and TM modes, $\mathrm{\mathbf{E}^*_{\mathrm{TE}}(\mathbf{r})}\times{\mathbf{E}_{\mathrm{TM}}(\mathbf{r})}$ lies in the \emph{x-y} plane and therefore can couple to the in-plane component along the $z$ axis of the magnon modes. For the YIG thin film, the Faraday rotation coefficient $\theta_{\mathrm{F}}$ $=$ 200$^{\circ}$/cm, $\lambda_{\mathrm{o}}$ $\approx$ 1520 nm, refractive index $n_{\mathrm{YIG}}$ $=$ 2.19, $\omega_{\mathrm{TE}} \approx \omega_{\mathrm{TM}} \approx 2\pi\times197.2$ THz, and the total length of the waveguide being 4.2 mm. From the simulations of the optical and magnon modes field distributions, and assuming near uniform field distributions along the waveguide longitudinal direction, we can calculate that the effective mode volume of magnon, optical TE, and optical TM modes being 1.47 $\times$ $10^{-13}$ $\mathrm{m^3}$, 7.25 $\times$ $10^{-14}$ $\mathrm{m^3}$, and 6.92 $\times$ $10^{-14}$ $\mathrm{m^3}$, respectively, showing that the optical TE and TM modes have very similar mode size and magnon mode is relatively less confined. 

The maximal value of the single photon coupling strength is then calculated to be $G_{\mathrm{TE,TM},\gamma} = 2\pi \times$ 22.7 Hz, assuming the uniform distribution of both optically-generation magnetic field and magnon field. The normalized spatial dependence of the magneto-optical coupling strength $|G(\mathrm{\mathbf{r}})|$ among optical TE, TM, and magnon modes at the cross-section view is plotted in the Fig. \ref{figures1}. As we can see, the YIG waveguide geometry supports the coexistence of optical and magnon modes with large overlap factor and small mode volume. Considering the spatial distributions of both optical and magnon modes along the longitudinal direction, the single photon coupling strength for the fundamental magnon mode is calculated as $G_{\mathrm{TE,TM},1} = 2\pi \times$ 8.15 Hz.

\section*{Appendix B. Device Fabrications}

The fabrication of the optomagnonic waveguides start with a (111)-oriented single crystalline YIG-on-GGG wafer, with the thickness of YIG thin film and GGG substrate being 5 $\mu$m and 500 $\mu$m, respectively. To mitigate the challenge of etching single crystal YIG while achieving smooth surface finish, we develop a lithography process combining both high-resolution dry etch and acid wet etch. We first define a hard mask for etching the 5-$\mu$m-wide top layer of the YIG waveguides. Here, a 250-nm-thick silicon dioxide (${\textrm{SiO}_{2}}$) is deposited via plasma-enhanced chemical vapor deposition (PECVD) as the hard mask. This oxide mask layer is patterned via electron-beam lithography (EBL) using the Ma-N 2403 resist followed by the chlorine-based reactive ion-etching (RIE). The YIG layer is then wet etched by 1.2\,$\mu$m using hot phosphoric acid at around 140 $^{\circ}$C. 

Next, to etch bottom slab layer, a thick hard mask (2-$\mu$m ${\textrm{SiO}_{2}}$) is redefined on the etched film using PECVD
and EBL patterning. Then, the patterned film is etched by another 5 $\mu$m into the GGG layer using hot phosphoric acid. Finally, a 6-$\mu$m-thick PECVD ${\textrm{SiO}_{2}}$ layer is covered on the chip surface to provide mechanical support and minimize the optical scattering loss at the device surface. The final device is diced and polished at both facets using diamond polishing paper to reduce the surface roughness. Finally, the chip facets are coated with 1-nm-thick titanium (Ti) and 9-nm-thick gold (Au) to form the reflection mirrors using ebeam evaporation.

\bibliography{ref}% Produces the bibliography via BibTeX.

\end{document}